\documentclass[%
 reprint,
superscriptaddress,
 amsmath,amssymb,
 aps,
]{revtex4-2}

\usepackage{graphicx}
\usepackage{dcolumn}
\usepackage{bm}

\usepackage{gensymb}

\usepackage{braket}
\usepackage{xcolor}
\usepackage{kotex}

\begin{document}

\preprint{APS/123-QED}

\title{Crucial role of magnetic dipole and anomalous field dependence in $\mathrm{NiPS_3}$}

\author{Do Hoon Kiem}
\affiliation{%
Department of Physics, Korea Advanced Institute of Science and Technology (KAIST), Daejeon 34141, Republic of Korea}%

\author{Muhammad Nauman}
\affiliation{%
Department of Physics, Prairie View A\&M University, Prairie View, Texas, USA 
}%
\affiliation{
Institute of Science and Technology Austria (ISTA), Klosterneuburg 3400, Austria
}

\author{Joonyoung Choi}
\affiliation{
Department of Physics, Kyungpook National University, Daegu 41566, Republic of Korea
}

\author{David Graf}
\affiliation{
National High Magnetic Field Laboratory, Tallahassee, FL 32310
}

\author{Junghyun Kim}
\affiliation{
Department of Physics and Astronomy, Seoul National University, Seoul 08826, Republic of Korea
}
\affiliation{
Center for Quantum Materials, Seoul National University, Seoul 08826, Republic of Korea
}

\author{Heejun Yang}
\affiliation{
Department of Physics and Astronomy, Seoul National University, Seoul 08826, Republic of Korea
}
\affiliation{
Center for Quantum Materials, Seoul National University, Seoul 08826, Republic of Korea
}

\author{Je-Geun Park}
\affiliation{
Department of Physics and Astronomy, Seoul National University, Seoul 08826, Republic of Korea
}
\affiliation{
Center for Quantum Materials, Seoul National University, Seoul 08826, Republic of Korea
}

\author{Younjung Jo}
\email{jophy@knu.ac.kr}
\affiliation{
Department of Physics, Kyungpook National University, Daegu 41566, Republic of Korea
}

\author{Myung Joon Han}
\email{mj.han@kaist.ac.kr}
\affiliation{%
Department of Physics, Korea Advanced Institute of Science and Technology (KAIST), Daejeon 34141, Republic of Korea}%

\begin{abstract}
$\mathrm{NiPS_3}$ has attracted considerable attention, yet the origin of its intriguing magnetic behaviors remains elusive due to limited insight into its low-energy anisotropy. Here, high-field torque magnetometry measurements are performed, revealing a previously unrecognized anisotropic spin response that reflects interaction terms beyond the standard XY or XXZ models. In combination with relativistic density functional theory calculations and magnetic force response theory, an effective low-energy spin Hamiltonian is constructed, elucidating not only the observed anomalous magnetic behavior but also other recent experiments. Remarkably, the magnetic dipole interaction and the associated spin-flop transition play a crucial role in driving the deviation from two-dimensional spin alignment, providing direct insight into the underlying anisotropic energy hierarchy. The generalized Hamiltonian can serve as the important fundamental ground for future study, and our approach adopted here will hopefully be relevant to many other related material systems.
\end{abstract}

\maketitle


Magnetic van der Waals (vdW) materials have emerged as fertile platforms for both fundamental physics and device applications \cite{burch2018magnetism,lee_ising-type_2016,huang2017layer,fei_two-dimensional_2018,wang2022magnetic}. $\mathrm{NiPS_3}$ is a particular representative case under active recent investigation \cite{kang2020coherent,wang2021spin,hwangbo2021highly,afanasiev2021controlling, belvin2021exciton,wang2022electronic,he2024magnetically,klaproth2023origin,scheie2023spin,jana2023magnon,kim2023rapid,lee2024imaging,wildes2022magnetic,ergeccen2022magnetically, sun2024dimensionality,song2024manipulation,wang2024unveiling,dirnberger2022spin,kim2023anisotropic,kim2019suppression} as a prototypical antiferromagnet with XY-like (or XXZ) spins, supported by susceptibility and thickness-dependent magnetization measurements \cite{kang2020coherent,sun2024dimensionality,kim2019suppression,joy1992magnetism,chandrasekharan1994magnetism}. 
Recently, a series of experiments report the intriguing many-body physics in $\mathrm{NiPS_3}$ \cite{kang2020coherent,wang2021spin,hwangbo2021highly,afanasiev2021controlling,belvin2021exciton,wang2022electronic,ergeccen2022magnetically,dirnberger2022spin,he2024magnetically}. In order to scrutinize the characteristic behaviors of this material, new theoretical models or pictures have been suggested such as Hund’s exciton \cite{he2024magnetically} and coherent many-body exciton \cite{kang2020coherent,wang2021spin,hwangbo2021highly} associated with Zhang-Rice triplet-singlet transition \cite{kang2020coherent,klaproth2023origin,song2024manipulation,li2024charge}.

It is quite certain that these observations are intrinsically attributed to magnetism, but remain not captured by the simple XY model. For example, the sudden change in photoluminescence (PL) peak polarization likely indicates higher-order anisotropy \cite{wang2021spin,jana2023magnon}. The critical exponent was found to be larger than that of XY model \cite{wildes2015magnetic}. Neutron measurements also suggest symmetry-breaking terms within the easy-plane \cite{scheie2023spin,wildes2022magnetic,wildes2015magnetic,lanccon2018magnetic}. Optical pump polarization is thought to be strongly associated with in-plane anisotropy \cite{afanasiev2021controlling,belvin2021exciton}, and resonant inelastic X-ray scattering (RIXS) data are interpreted as having a correlated excitonic origin \cite{kang2020coherent,he2024magnetically}. While these experiments demonstrate how the great advancements of modern measurement techniques unveil otherwise hidden critical details of the low-energy excitations and responses, their comprehensive understanding remains far from clear and is still under active debate \cite{kang2020coherent,wang2021spin,he2024magnetically,klaproth2023origin}. A full account of anisotropic interactions beyond the XY model is required.

In principle, such an effective magnetic Hamiltonian beyond the XY model can be constructed by incorporating relativistic corrections of order $1/(mc)^2$ as derived from quantum electrodynamics \cite{itoh1965derivation}. For real materials, however, constructing a Hamiltonian that includes terms such as Kitaev-$\Gamma$, Dzyaloshinskii–Moriya (DM), and single-ion anisotropy, and clarifying their roles, remains highly challenging \cite{pesin2010mott,rau2014generic, dzyaloshinsky1958thermodynamic,moriya1960anisotropic}. This requires precise determination of energy scales typically in the $\mu$eV to sub-meV range, which remains demanding even for state-of-the-art theoretical and experimental methods.

To unravel the anisotropic magnetic profile in $\mathrm{NiPS_3}$, we conducted torque magnetometry together with relativistic density functional theory (DFT) calculations. The high-field torque response ($H\geq9$ T) reveals an intriguing anomaly. Furthermore, complete three-dimensional (3D) understanding of torque data, combined with the results of relativistic DFT and magnetic force calculation, provides the detailed anisotropic interaction profile. From this we established a generalized spin Hamiltonian incorporating terms down to the $\mu$eV scale, which consistently explains both the previously reported many-body magneto-optical features \cite{wang2021spin,jana2023magnon,song2024manipulation,wang2024unveiling} and torque anomaly associated with an out-of-plane spin-flop transition. Remarkably, the dipolar interaction plays an essential role, being directly responsible for the characteristic torque and PL behaviors by favoring a specific energetic order. Our results establish a unified framework to interpret recent experiments and guide future applications.

\begin{figure}
    \centering
    \includegraphics[width=1.0\linewidth]{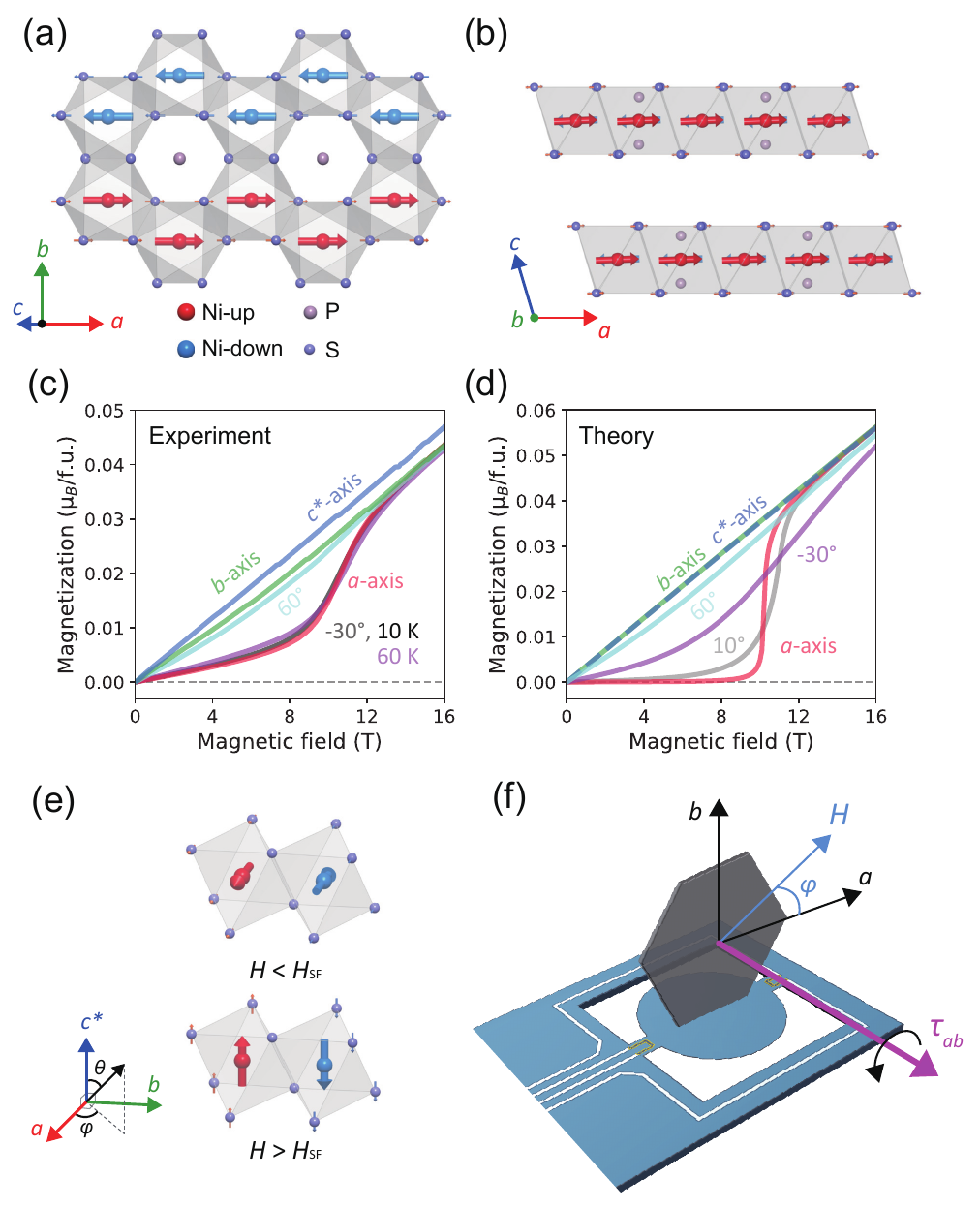}
    \caption{
    (a) Top and (b) side view of crystal structure and the ground state spin configuration of $\mathrm{NiPS_3}$. The red and blue arrows indicate the up and down spins, respectively. 
    (c) The measured magnetization using VSM up to $H = 16$ T. The measurement temperature was $T=10$ K for the field directions of $H_{\parallel a}$ ($\theta=90\degree$,$\varphi=0\degree$), $b$ ($\theta=90\degree$,$\varphi=90\degree$),$c^*$-axes ($\theta=0\degree$), $\varphi=60\degree$ and $-30\degree$ ($\theta=90\degree$). For $\varphi=-30\degree$, an additional measurement was conducted also at $T=60$ K. 
    (d) The corresponding calculation results using the computed parameters (see Fig.~\ref{fig:figure3}).
    (e) The schematic figure for the spin orderings under the applied field $H$ above and below the spin-flop field $H_\mathrm{SF}$. At low field, spins align along the $a$-axis ($\theta=90\degree$, $\varphi$=$0\degree$). 
    (f) Schematic picture of torque magnetometer. The sample $ab$-plane is depicted by the gray hexagonal plane within which the applied magnetic field $H$ varies with angle $\varphi$ (blue arrow). The out-of-plane torque ($\tau_{ab}$; violet arrow) is measured by the circuit. For more details, see Fig.~S2.}
    \label{fig:figure1}
\end{figure}

$\mathrm{NiPS_3}$ consists of edge-shared $\mathrm{NiS_6}$ octahedra forming a honeycomb lattice and exhibits a zigzag antiferromagnetic (AFM) order below $T_\mathrm{N}=155$~K with $C_{2/m}$ symmetry due to the stacking as shown in Fig.~\ref{fig:figure1}(a,b) \cite{kang2020coherent,he2024magnetically,kim2018charge,kim2019mott,lanccon2018magnetic,wildes2015magnetic,kim2019suppression}. Previous magnetization studies were limited to $H \leq 9$ T \cite{wildes2015magnetic,lanccon2018magnetic,selter2021crystal,shemerliuk_tuning_2021}. Here, we extended the measurements up to $H = 16$ T, as summarized in Fig.~\ref{fig:figure1}(c) (see Fig.~\ref{fig:figure1}(e) for the angle convention). At low fields $H \leq 9$ T, the magnetization increases linearly and shows larger values along the $c^*$-axis than along the $b$- and $a$-axes ($M_{\parallel a} < M_{\parallel b} \leq M_{\parallel c^*}$), consistent with previous reports \cite{wildes2015magnetic,lanccon2018magnetic,selter2021crystal,shemerliuk_tuning_2021}. The in-plane magnetization exhibits easy-axis anisotropy, already deviating from the ideal XY behavior \cite{joy1992magnetism,chandrasekharan1994magnetism}, in agreement with the susceptibility data (Fig.~S4). At high fields around $H \approx 10$ T, a characteristic up-turn feature appears, unreported in earlier studies, and is observed only when the field is applied nearly along the $a$-axis. Above $H \geq 12$ T, the magnetization curves for the $a$- and $b$-axes eventually merge. Notably, this critical field coincides with the reported $H \approx 10.5$ T at which a sudden change in PL polarization occurs \cite{wang2021spin}. Below we will show that these two experiments share the same origin. 

\begin{figure}
    \centering
    \includegraphics[width=1.0\linewidth]{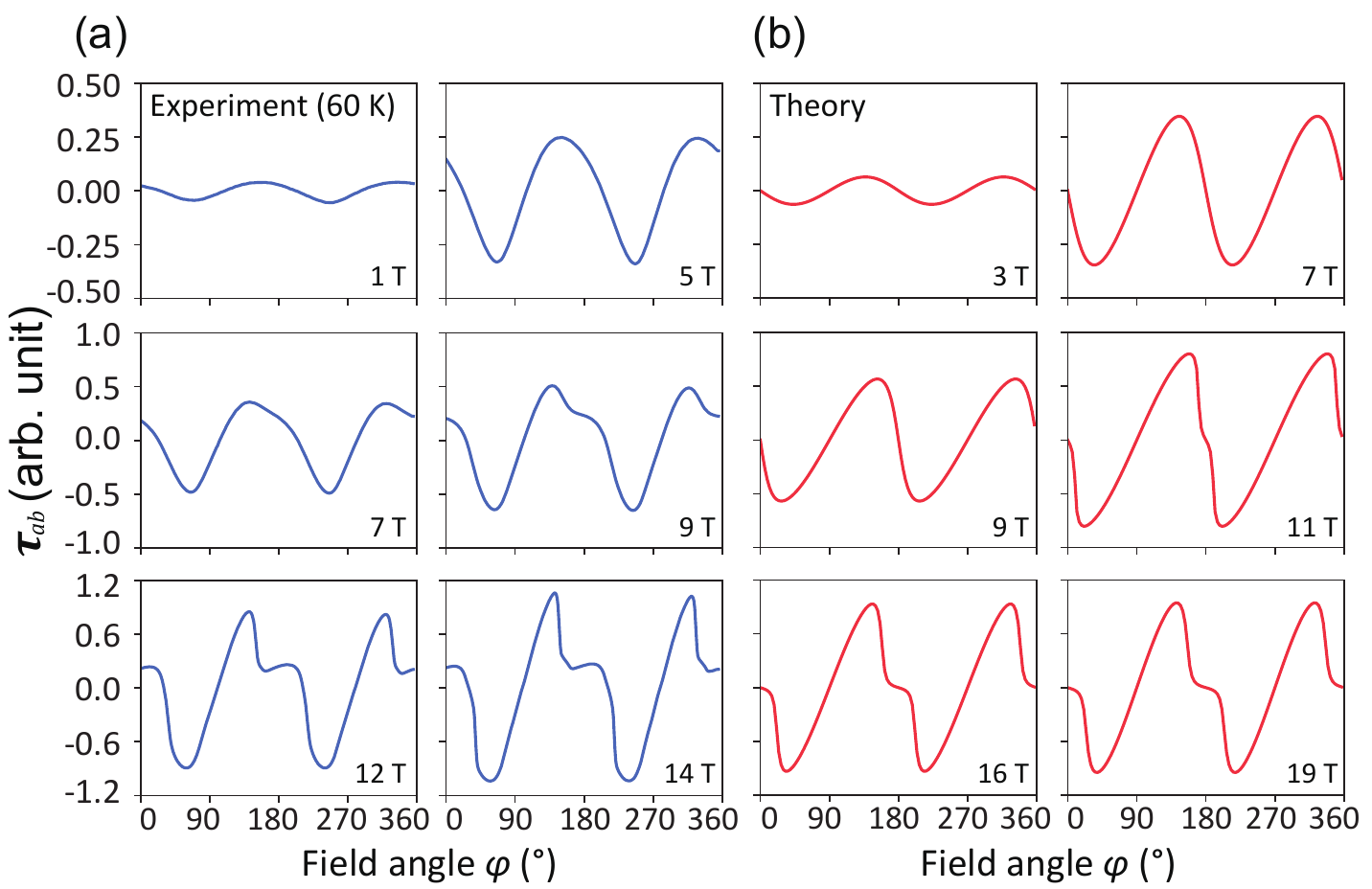}
    \caption{
    (a) The measured torque $\tau_{ab}$ obtained at $T = 60$ K and with varying magnetic fields. The in-plane field direction $\varphi$ is depicted in Fig.~\ref{fig:figure1}(f) (b) The corresponding calculation results of $\tau_{ab}$ as obtained from the constructed spin Hamiltonian. }
    \label{fig:figure2}
\end{figure}

To map out the detailed magnetic Hamiltonian profile, we performed the torque magnetometry measurements as depicted in Fig.~\ref{fig:figure1}(f). Below $T_\mathrm{N}$, the torque is expressed as $\vec{\tau}=(\overleftrightarrow{\chi}\cdot\vec{H})\times\vec{H}$. Owing to the mirror symmetry, the off-diagonal components of susceptibility ($\chi_{ab}$, $\chi_{bc}$) vanish \cite{nauman2021complete}. Under an in-plane rotating field $\vec{H}=H(\cos{\varphi},\sin{\varphi},0)$, the torque thus simplifies to describe the angular dependence in the linear-response regime: 
\begin{equation}
    \tau_{ab}=\frac{1}{2}\left(\chi_{aa}-\chi_{bb}\right)H^2 \sin{2\varphi}.
    \label{eq1:simpletorque}
\end{equation}
Figure~\ref{fig:figure2}(a) shows that the measured torque $\tau_{ab}$ 
indeed follows the $\sin 2\varphi$ dependence well in the low field $H\leq5$ T. This simple trigonometric form of data already shows the need of in-plane magnetic anisotropy terms beyond the ideal XY spin model, as discussed above with magnetization data. 

An intriguing new feature starts to emerge as the field strength increases. As shown in Fig.~\ref{fig:figure2}(a), a characteristic ``shoulder'' develops at $\varphi \approx 180\degree$ when the field strength is greater than $H \approx 7$--$9~\text{T}$. It eventually evolves into a well-distinguished second peak at $H \geq 12~\text{T}$. Obviously, this new shape cannot be captured by $\sin 2\varphi$.
Hypothesizing that it originates from an intrinsic material characteristic, we consider a general form of spin Hamiltonian as follows:
\begin{equation}
\label{eq:ham_total}
\begin{aligned}
\mathcal{H} 
  =& -\sum_{i \neq j} \hat{{S}}_i \cdot \mathbf{J}_{ij} \cdot \hat{{S}}_j
     + \sum_i \hat{{S}}_i \cdot \mathbf{K}_i \cdot \hat{{S}}_i \\
     &+ \sum_{i,j} D(\hat{{S}}_i, \hat{{S}}_j)
     - \sum_i \vec{m}_i \cdot \vec{H}.
\end{aligned}
\end{equation}
Here, $\hat{S}_i$ and $\hat{S}_j$ are the unit spin vectors at the local sites $i$ and $j$, respectively. $\mathbf{J}_{ij}$ denotes the exchange coupling tensor that includes both isotropic and anisotropic exchange interactions such as Heisenberg, DM, and Kitaev-$\Gamma$ interactions~\cite{rau2014generic,pesin2010mott,dzyaloshinsky1958thermodynamic,moriya1960anisotropic,szilva2023quantitative,udvardi2003first}. 
The second term represents the single-ion anisotropy, expressed as $\sum_i K_i(\hat{{e}}_i\cdot\hat{{S}}_i)^2$ for a uniaxial case with strength $K_i$ along the local axis $\hat{{e}}_i$. Both exchange and single-ion anisotropies originate from spin–orbit coupling, and therefore one needs to resort to relativistic DFT. In the current study, they were obtained respectively from the magnetic force linear response theory (MFT) and the DFT total energy calculation. The third term $D(\hat{{S}}_i, \hat{{S}}_j)$ corresponds to the magnetic dipole--dipole interaction and is given by
\begin{equation}
D(\hat{{S}}_i, \hat{{S}}_j) = 
\frac{\mu_0 m_i m_j}{4\pi r_{ij}^3} 
\Big[ \hat{{S}}_i \cdot \hat{{S}}_j 
- 3(\hat{{S}}_i \cdot \hat{{r}}_{ij})(\hat{{S}}_j \cdot \hat{{r}}_{ij}) \Big],
\end{equation}
where $\mu_0$ is the vacuum permeability and $m_i, m_j$ denote the magnetic moments at each site.
The torque data can be compared by taking the external field into account; $\mathcal{H}_{\mathrm{ext}} = -\sum_i \vec{m}_i \cdot \vec{H}$. Once the parameters of each term are given, the corresponding total energy 
$E(\{\hat{{S}}_i\}; \vec{H})$ depends on the moment directions, and one can obtain the torque by minimizing the energy:
\begin{equation}
\vec{\tau} = \vec{M}_{\mathrm{total}} \times \vec{H} 
= \sum_i \vec{m}_i \times \vec{H},
\label{eq:torque}
\end{equation}
where $\vec{M}_{\mathrm{total}}$ is the induced magnetization. 
Below, we try to construct the Hamiltonian and systematically investigate each term. Finally, Equation~\eqref{eq:torque} gives rise to the result shown in Fig.~\ref{fig:figure2}(b) in excellent agreement with the experiment.

\begin{figure*}
    \centering
    \includegraphics[width=0.8\linewidth]{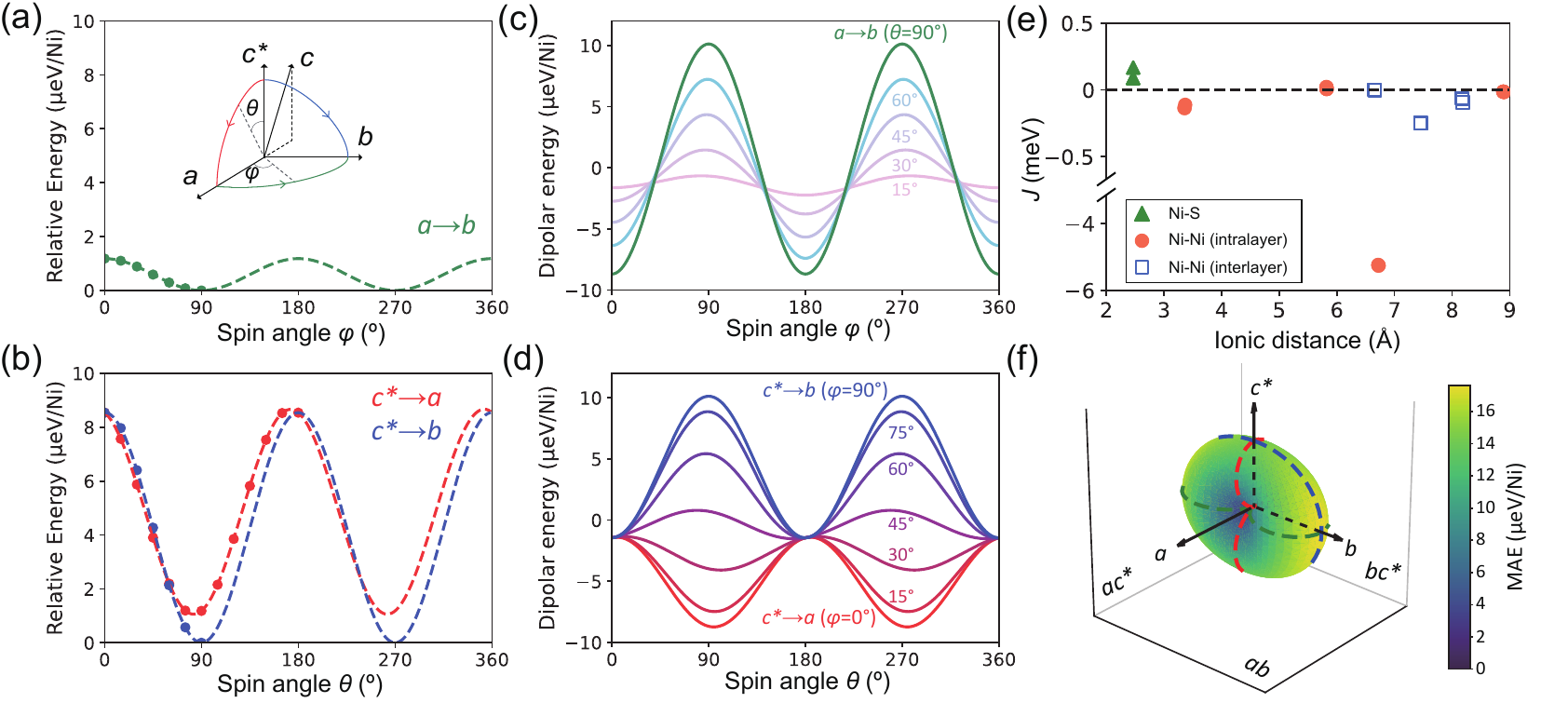}
    \caption{
    (a, b) The calculated magneto-crystalline energy by varying spin-rotation angles $\varphi$ and $\theta$. The dashed lines are extrapolated from the calculated relative energies (filled circles). The inset of (a) depicts the rotation paths and angles. For $\theta$ dependence, we considered both $c^{*}a$- (red) and $c^{*}b$-plane (blue) spin rotations. (c,d) The calculated dipole interaction energies for (c) in-plane and (d) out-of-plane rotations. Different colors represent the different rotation angles. The rotation angles $\varphi$ and $\theta$ are the same as in the inset of (a). (e) The MFT calculation results of isotropic Heisenberg exchange coupling $J$ as a function of interatomic distance. By convention, $J > 0$ corresponds to ferromagnetic (FM) interactions, while $J < 0$ corresponds to AFM interactions. Filled triangles (green), circles (orange), and empty squares (light blue) denote the Ni-S, intralayer Ni-Ni, and interlayer Ni-Ni couplings, respectively. 
    (f) The 3-dimensional visualization of the calculated total magnetic anisotropy energy, including magneto-crystalline energy and dipole interaction. At each point of the total energy surface, the distance from the origin and the color correspond to the anisotropic energy for the given direction. The three dashed-line curves show the energy profiles at three different planes of $bc^{*}$, $ac^{*}$, and $ab$. For more quantitative presentations, see Fig.~S8. 
    }
    \label{fig:figure3}
\end{figure*}

We now examine the contribution of each term. The first term of Eq.~\ref{eq:ham_total} is calculated using MFT combined with relativistic non-collinear DFT \cite{szilva2023quantitative,kashid2014dzyaloshinskii,borisov2021heisenberg}. Previous studies treated the exchange couplings as isotropic Heisenberg ($J_{xx}=J_{yy}=J_{zz}$) as a common assumption \cite{scheie2023spin,wildes2022magnetic,lanccon2018magnetic,mi2022variation}.
In contrast, our non-collinear MFT makes no a priori assumptions and directly evaluates the full exchange tensor $\mathbf{J}_{ij}$ for each Ni–Ni and Ni–S interaction. For comparison with earlier works, we present the isotropic average $J=(J_{xx}+J_{yy}+J_{zz})/3$, summarized in Fig.~\ref{fig:figure3}(e) (see Table~S1).
The third-nearest intra-layer Ni–Ni interaction at 6.7 \AA (see also in Fig.~S6) is found to be the dominant coupling, stabilizing the known zigzag ground state~\cite{wildes2015magnetic,lanccon2018magnetic,mi2022variation,kim2021magnetic}. In addition, our calculations reveal previously unexplored Ni–S interactions and long-range profiles.
Off-diagonal components of $\mathbf{J}_{ij}$, including antisymmetric DM and symmetric Kitaev-$\Gamma$ terms, are also finite with a magnitude of approximately 0.01 meV (Fig.~S5).

The magneto-crystalline anisotropy (MCA), corresponding to the first two terms of Eq.~\ref{eq:ham_total}, is directly evaluated from relativistic DFT total energies. The results for rotating spin directions are shown in Fig.~\ref{fig:figure3}(a,b). As expected, the MCA is small, with a maximum of 8.5 $\mu$eV/Ni for the $c^{*}a$ rotation. The total energy profile identifies the $ab$-plane as the easy plane and the $c$-axis as the hard axis. A subtle feature appears for the $c^{*}a$ rotation, where the energy minimum ($\theta\approx82\degree$) slightly deviates from $\theta=90\degree$, reflecting the broken mirror and residual $C_2$ symmetry \cite{nauman2021complete}. The $ab$- and $bc^{*}$-plane retain mirror symmetry with the sinusoidal dependence. Despite the small energy scale, the DFT results capture the essential symmetry features of the anisotropy, supporting their reliability.

Here, we pay special attention to the energetics revealed by MCA. The calculated energy profile distinguishes the $a$-axis from the $b$-axis as expected from the low-field torque data (Fig.~\ref{fig:figure2}(a)). However, the high-field features exhibiting ‘shoulders’ cannot be reconciled with the energetic order $E_{\parallel b}<E_{\parallel a}<E_{\parallel c^*}$ (Fig.~\ref{fig:figure3}(a,b)). The torque computed from the corresponding Hamiltonian (first two terms in Eq.~\ref{eq:ham_total}) also fails to reproduce the experimental profile (Fig.~S6(a)), highlighting the missing contribution of the dipolar interaction.


Magnetic dipole–dipole interaction is typically very weak in real materials and often neglected \cite{marder2010condensed,ashcroft1976solid,blundell2001magnetism}, though they can contribute to anisotropy in large moment systems \cite{lagendijk1972caloric,jensen1991rare}. Incorporating this effect quantum-mechanically into the DFT scheme is known to be challenging despite recent progress \cite{pellegrini2020density,jansen1999magnetic}. In $\mathrm{NiPS_3}$, with $L=0$, we include this term classically using the Ni ($\sim$1.5 $\mu_B$) and S ($\sim$0.1 $\mu_B$) spin moments obtained from relativistic DFT. Figures~\ref{fig:figure3}(c,d) show the calculated dipolar energy with respect to the in-plane and out-of-plane spin rotation, respectively. Note that the energy minimum is found at $\theta\approx90\degree$ (in $ca$-plane), consistent with the experimental easy-axis \cite{lanccon2016magnetic}. This dipolar interaction yields $E_{\parallel a} < E_{\parallel c^{*}}<E_{\parallel b}$. 


In Fig.~\ref{fig:figure3}(f), we summarize the magnetic total energy profile as a 3D contour map encompassing all anisotropy terms in Eq.~\ref{eq:ham_total}. It is noted that the overall energetic order is now $E_{\parallel a}<E_{\parallel c^{*}}<E_{\parallel b}$ as indicated by the three dashed lines on the surface and also in Fig.~S8. Importantly, this order matches the dipolar energy trend shown in Fig.~\ref{fig:figure3}(c,d), demonstrating that the dipolar interaction plays a pivotal role in determining the magnetic anisotropy of this material.


Using this final form of Hamiltonian, we recalculate torque as a function of field direction and strength. As mentioned above, the result shown in Fig.~\ref{fig:figure2}(b) is in excellent agreement with the experiment in Fig.~\ref{fig:figure3}(a). Hereby, we conclude that the dipole interaction is the key intrinsic source of magnetic anisotropy responsible for the observed high-field torque data. It is remarkable that the dipolar contribution plays such an important role that a qualitatively different signature is detected in the experiment. 


Interestingly, the observed energy hierarchy, $E_{\parallel a}<E_{\parallel c^{*}}<E_{\parallel b}$, deviates from the easy-plane expectations of the XXZ model ($E_{\parallel a}\leq E_{\parallel b}<E_{\parallel c^{*}}$). XY or XXZ models assume spins confined within the $ab$-plane, allowing only weak in-plane anisotropy. In contrast, our results indicate that the spin degrees of freedom can no longer be considered strictly in-plane, as will become more evident in subsequent discussion of the spin-flop transition. Model simulations, performed with various sets of arbitrarily chosen parameters, consistently reproduce the qualitative shoulder feature as long as the energy hierarchy $E_{\parallel a}<E_{\parallel c^{*}}<E_{\parallel b}$ is retained (Fig.~S9). In realistic parameter regimes, the dipolar interaction stabilizes this feature. 


\begin{figure*}
    \centering
    \includegraphics[width=1.0\linewidth]{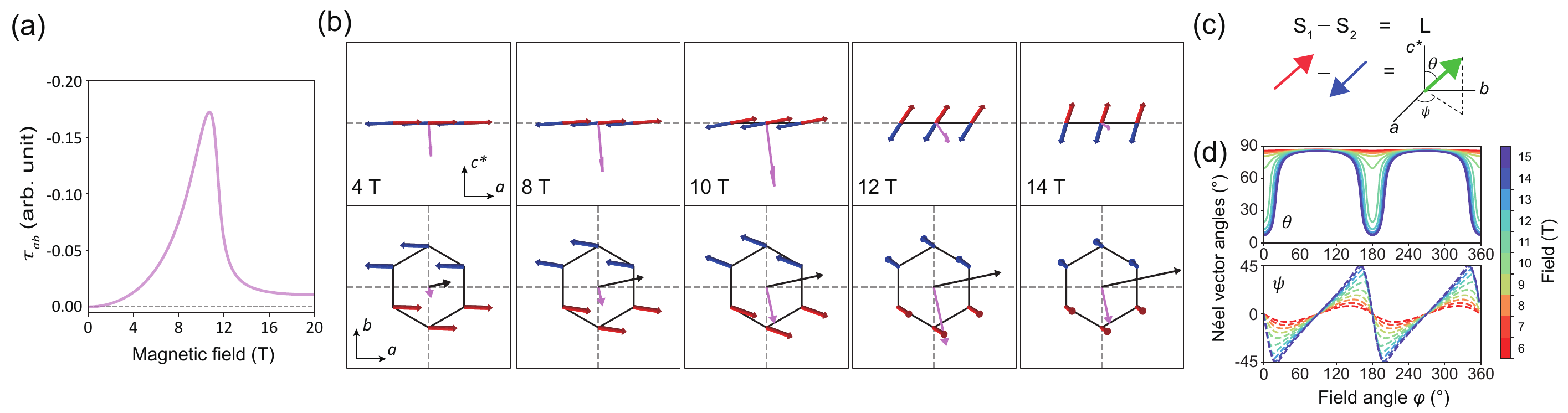}
    \caption{
    (a) The calculated torque $\tau_{ab}$ as a function of a magnetic field. The peak developed at $H$ = 10.2 T indicates the spin-flop. (b) Snapshot pictures showing the spin-flop transition and the related changes around the critical field strength. The top and bottom panels depict the side and top views. The black honeycomb indicates the honeycomb lattice of Ni. The red/blue arrows depict the up/down spin moments of Ni ions. The black and violet arrows show the magnetic field and the torque vectors, respectively. (c) The definition of the polar and azimuthal angle of the Néel vector. (d) The calculated Néel vector angle changes ($\theta$ and $\psi$) by varying the in-plane field angle ($\varphi$). Note the sudden changes of both $\theta$ and $\psi$ at around the spin-flop critical field strength of 10--11 T. 
    }
    \label{fig:figure4}
\end{figure*}

Let us now further scrutinize experimental data based on the Hamiltonian constructed above. Figure \ref{fig:figure4}(a) shows the calculated torque as a function of field applied nearly along the $a$-axis ($\varphi=12\degree$). While $\tau_{ab}$ gradually increases along with the field strength, it becomes notably steeper at around $H\approx 8$ T up to the critical field $H_\mathrm{SF}$ = 10.2 T, and suddenly drops at $H \geq H_{\mathrm{SF}}$, indicating a spin-flop transition. It is clearly demonstrated with  snapshot pictures in Fig.~\ref{fig:figure4}(b) with moments (red and blue arrows) and torque (violet arrow). Below $H_\mathrm{SF}$, spins lie within the $ab$-plane (i.e., $\theta\approx90\degree$; see also Fig.~\ref{fig:figure1}(c)). As the field strength increases and reaches to $H\approx H_{\mathrm{SF}}$, they stand up, while maintaining the same AFM spin order, and make a sizable change in directions (12 T and 14 T). At the same time, the out-of-plane torque component is significantly reduced, consistent with the measured torque and responsible for the characteristic 'shoulder' features in Fig.~\ref{fig:figure2}. For the more clear presentation, see Movie S1.

Finally, we consider the anisotropic PL response recently reported by Wang et al.~\cite{wang2021spin}, which has been proposed for coherent many-body exciton manipulation, opto-spintronic devices, and magnetic quantum information \cite{kang2020coherent,wang2021spin,afanasiev2021controlling,belvin2021exciton,song2024manipulation,wang2024unveiling,liu20192d}. However, understanding this phenomenon remains unclear due to the lack of detailed theoretical analysis. It is clear that the observation indicates anisotropy beyond the conventional XY-type interactions \cite{wang2021spin,afanasiev2021controlling,belvin2021exciton,scheie2023spin,jana2023magnon}. Interestingly, the critical field for the anomalous PL response ($H_\mathrm{PL}=10.5$ T~\cite{wang_defect-mediated_2021,jana2023magnon,song2024manipulation,wang2024unveiling}) is close to those observed in our magnetization ($H_\mathrm{MM}\sim10$ T) and torque ($H_\tau=$7–9 T) measurements.


Based on the Hamiltonian, we calculated the field-dependent response of the Néel vector, $\vec{L}=\vec{S}_1-\vec{S}_2$ (Fig.~\ref{fig:figure4}(c)), a key quantity in the previous reports~\cite{wang2021spin,afanasiev2021controlling}.
The Néel vector carries crucial information for understanding the excitonic behavior. The spin-flop transition involves a sudden reorientation of $\vec{L}$.
As shown in Fig.~\ref{fig:figure4}(d), the out-of-plane angle $\theta$ rapidly changes from $90\degree$ to $0\degree$ around $H$=10–11 T, while the in-plane angle $\psi$ evolves continuously up to $45\degree$. This reorientation of $\mathbf{L}$ accounts for the sudden change in PL, with the dipolar interaction playing a crucial role in shaping this characteristic response.


We note that spin-flop is the key to understanding not only the torque anomaly but also the high-field magnetization behavior presented in Fig.~\ref{fig:figure1}(d). Figure~\ref{fig:figure1}(e) presents the calculated magnetization based on the generalized Hamiltonian. It is reasonably well compared with experimental data by capturing the rapid up-turn along the $a$-axis and the linear behavior (no up-turn) along the $b$- and $c$-axes. The analysis shows that applying the high enough field along the $a$-axis flops the spins to align along the $c$-axis (as in Fig.~\ref{fig:figure4}(b)), causing the rapid increase in magnetization.

Some discrepancies remain between experiment and calculation. The calculation gives the identical $b$- and $c^{*}$-axis magnetizations, whereas experimentally the $c$-axis magnetization is slightly larger. Similar to its sister compound $\mathrm{FePS_3}$ \cite{lanccon2016magnetic,murayama2016crystallographic}, the weak vdW interaction allows the formation of $120\degree$-rotated twinned domains along the $c^{*}$-axis \cite{scheie2023spin}. As shown in the Supplemental Materials, coexistence of such twinned domains reproduces not only the weak linear increase of magnetization along the $a$-axis for $H<H_\mathrm{SF}$, resulting in $M_{\parallel b} < M_{\parallel c^{*}}$, but also the finite torque values at $\varphi=0\degree$ and $180\degree$, consistent with experiment~(see Figs.~S10 and S11).

To unravel the unusual magnetic behaviors of the vdW magnet $\mathrm{NiPS_3}$, we combined torque magnetometry with first-principles calculations. The measured torque reveals anomalous high-field patterns. We analyzed these results by establishing the low-energy anisotropic spin Hamiltonian, incorporating Heisenberg, Kitaev-$\Gamma$, DM, single-ion anisotropy, and dipole interactions. The specific energetic order $E_{\parallel a}<E_{\parallel c*}<E_{\parallel b}$ is crucial for explaining the characteristic shoulder pattern in the torque data, with the dipolar interaction playing a key role in stabilizing it. Furthermore, we identify the spin-flop transition within this energetic hierarchy as the unifying mechanism underlying multiple anomalous responses, including high-field magnetization as well as torque and the anisotropic PL reported previously. Our findings provide key insights into understanding magnetic anisotropy in vdW magnets $\mathrm{NiPS_3}$, highlighting the critical role of dipole interactions in shaping high-field responses. Given the fundamental importance of these interactions, our results offer a robust theoretical framework for understanding magnetic anisotropy, guiding future studies and potential applications to vdW magnets.


\begin{acknowledgments}
M. J. H. was supported by a National Research Foundation of Korea (NRF) grant funded by the Korean government (MSIT) (Grant No. RS-2025-00559042). This work was supported by the Samsung Science \& Technology Foundation (Grant No. SSTF-BA2101-05). Y. J. was supported by a National Research Foundation of Korea (NRF) grant (RS-2024-00337640, RS-2023-00301914). A portion of this work was performed at the National High Magnetic Field Laboratory, which is supported by National Science Foundation Cooperative Agreement No. DMR-2128556 and the State of Florida. This work was also supported by the Leading Researcher Program of the National Research Foundation of Korea (Grant No. RS-2020-NR049405).

\end{acknowledgments}

\appendix


\bibliography{reference}

\end{document}